\documentclass[twocolumn,preprintnumbers,superscriptaddress,amsmath,amssymb,showpacs,showkeys]{revtex4}
\usepackage{color}
\usepackage{graphics}
\usepackage{dcolumn}
\usepackage{bm}

\renewcommand\textfraction 0
\renewcommand\topfraction 1
\renewcommand\bottomfraction 1
\begin{document} 

\preprint{Applied Physics A (2004) accepted}

\title{Electronic mechanism of ion expulsion\\ under UV nanosecond laser excitation of silicon:\\ Experiment and modeling}
\author{Wladimir \surname{Marine}}
\email{marine@crmcn.univ-mrs.fr}
\affiliation{Groupe de Physique des \'{E}tats Condens\'{e}s (GPEC), UMRS 6631 CNRS,
Universit\'{e} de la M\'{e}diterran\'{e}e, Case 901,163 Avenue de Luminy, 13288 Marseille Cedex 9, France}
\author{Nadezhda M. \surname{Bulgakova}}
\email{nbul@itp.nsc.ru}
\affiliation{Institute of Thermophysics, Prospect Lavrentyev 1, 630090 Novosibirsk, Russia}
\author{Lionel \surname{Patrone}}
\affiliation{Groupe de Physique des \'{E}tats Condens\'{e}s (GPEC), UMRS 6631 CNRS,
Universit\'{e} de la M\'{e}diterran\'{e}e, Case 901,163 Avenue de Luminy, 13288 Marseille Cedex 9, France}
\author{Igor \surname{Ozerov}}
\email{ozerov@crmcn.univ-mrs.fr}
\affiliation{Groupe de Physique des \'{E}tats Condens\'{e}s (GPEC), UMRS 6631 CNRS,
Universit\'{e} de la M\'{e}diterran\'{e}e, Case 901,163 Avenue de Luminy, 13288 Marseille Cedex 9, France}
 
\date{\today}


\begin{abstract}
We present experimental and modeling studies of UV 
nanosecond pulsed laser desorption and ablation of (111) bulk silicon. The 
results involve a new approach to the analysis of plume formation dynamics 
under high-energy photon irradiation of the semiconductor surface. 
Non-thermal, photo-induced desorption has been observed at low laser 
fluence, well below the melting threshold. Under ablation conditions, the 
non-thermal ions have also a high concentration. The origin of these ions is 
discussed on the basis of electronic excitation of Si surface states 
associated with the Coulomb explosion mechanism. We present a model 
describing dynamics of silicon target excitation, heating and charge-carrier 
transport.

\end{abstract}
\maketitle

\section{Introduction}

Laser ablation technique has been widely used for thin film deposition and 
nanoparticles synthesis. For these applications, understanding of dynamics 
and mechanisms of particle desorption and ablation from the irradiated 
surfaces is of crucial importance. For nanosecond laser pulses, it is 
generally accepted that the ablation mechanism of normal vaporization gives 
way to phase explosion with increasing laser fluence \cite{1}. At low laser 
fluences near ablation threshold, electronic mechanism of high-energy ion 
emission has been proven to play a role in initiating the ablation process, 
mainly for dielectrics and semiconductors \cite{2,3}. For ultrashort laser 
pulses, the electronic mechanisms of desorption \cite{4} and ablation \cite{5,6} for 
both dielectric and semiconductor targets are studied extensively, whereas 
for nanosecond pulses there is a lack of both the experimental and 
theoretical analyses. 

In this paper, we present the results of the experimental and theoretical 
studies of ion ejection from the Si targets induced by UV nanosecond laser 
pulses in a wide range of laser fluences in the regimes from those well 
below melting threshold up to developed ablation. Non-thermal ions have been 
detected at very low laser fluence. The theoretical study of ion ejection 
involves an analytical analysis and numerical modeling of charge transport 
in the laser-irradiated target. We attribute the high-energy ion emission to 
the generation of a strong electric field due to the electron photoemission 
process (the Coulomb explosion mechanism). 

\section{Experimental}

Clean (111) Si surface was passivated by hydrogen in a usual chemical way 
\cite{7}. Ablation was carried out in a high vacuum chamber (pressure $< 
10^{-8}$ Torr) at an incidence angle of 45\r{ } using an ArF 
excimer laser ($h\nu  = 6.4 $ eV, 15 ns pulse duration at FWHM). The set of masks 
was used to select the homogeneous part of the laser beam. The laser spot 
size on the target and laser fluence ($F_{0})$ were 0.5 mm$^{2}$ and $0.01 - 1 $ J/cm
$^{2}$, respectively. The target was rotated/translated during 
measurements to avoid cratering. The expansion dynamics and origin of the 
desorbed particles were analyzed by reflectron time-of-flight (TOF) mass 
spectrometer.

\begin{figure}[!]
\includegraphics{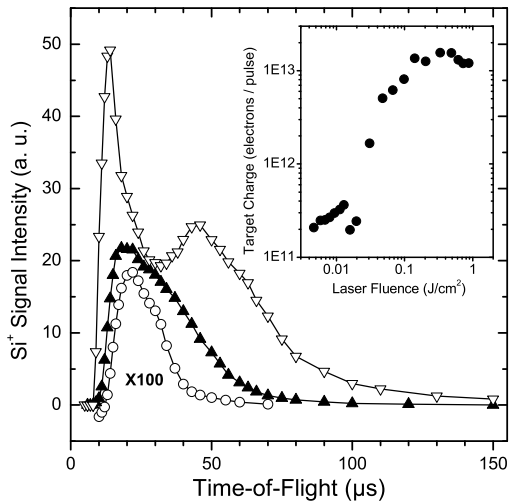} \caption{\label{fig1}
Time-of-flight spectra of Si$^{ + }$ ions for different 
laser fluences: \\open circles -- 0.26 J/cm$^{2}$ intensity (multiplied by 
100); \\dark triangles -- 0.42 J/cm$^{2 }$; \\open triangles -- 1.02 J/cm$^{2}$.\\ 
The target - mass spectrometer distance was 126 mm.\\ Inset shows charge of 
silicon target versus laser fluence.}
\end{figure}

\begin{figure}[!]
\includegraphics{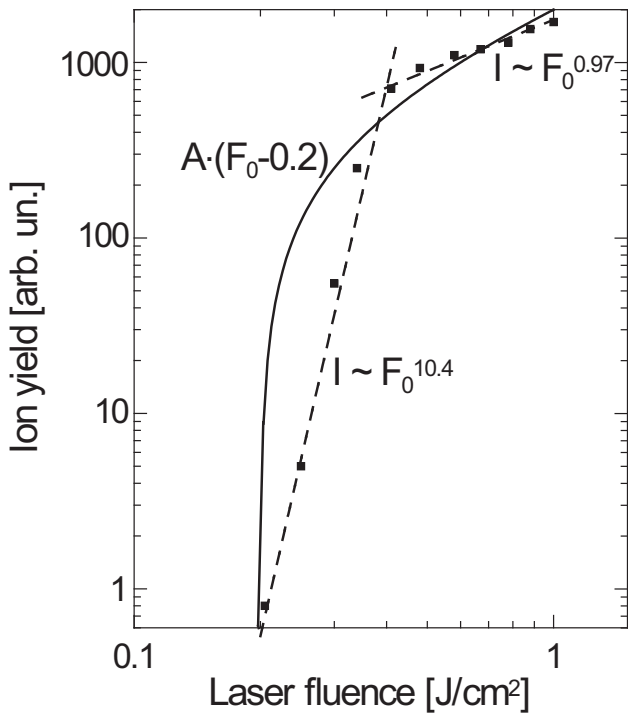} \caption{\label{fig2}
Experimental data on Si$^{ + }$ yield vs. laser fluence. 
0.2 J/cm$^{2}$ corresponds to the ion desorption threshold. Up to $\sim $0.4 
J/cm$^{2}$ that is around melting threshold, ion yield behaves as 
$F_{0}^{10.4}$ whereas at higher fluence dependence changes to 
$F_{0}^{0.97}$. Solid curve corresponds to Eq. (\ref{eq6}) with $F_{0th}$ = 0.2 
J/cm$^{2}$.}
\end{figure}

The most abundant desorbed species, Si$^{ + }$, are observed already at 
$F_{0} \sim $ 0.2 J/cm$^{2}$, well below melting threshold determined by 
time-resolved reflectivity measurements to be $\sim 0.4 $ J/cm$^{2} $ (see \cite{8} 
and references therein) that is consistent with the other studies \cite{9,10}. 
Neutral monatomic Si particles are detected only at $F_{0} > 0.8$ J/cm$^{2}$. 
Attempts to detect charged or neutral clusters, excepting silicon ions 
dimmer \cite{11}, have been unsuccessful. Figure \ref{fig1} shows the typical TOF spectra 
of Si$^{ + }$ at different fluences. At low laser fluence (0.26 J/cm$^{2 
}$on the Fig. \ref{fig1}, the TOF spectrum consists of an only population of Si$^{ + 
}$ ions. The spatial distribution of this kind of ions is narrow, 
strongly peaked relatively to the target normal and, up to the melting 
threshold, the kinetic energy ($\sim 4.8 - 5 $ eV) is a weak function of laser 
fluence that is typical for non-thermal desorption/ablation \cite{3,12}.

Increasing laser fluence above the melting threshold leads to broadening of 
Si$^{ + }$ TOF distribution (see Fig.~\ref{fig1}, fluence 0.42 J/cm$^{2})$ and to 
formation (from $0.7 - 0.8 $ J/cm$^{2})$ of a well-pronounced low-energy 
distribution \cite{13}. The appearance of the low-energy Si$^{ + }$ concurs with 
the development of Si surface melting. The intensity of this population 
rapidly increases with fluence and becomes dominant at $F_{0} > 1$ J/cm$^{2}$. 
The abundance of Si$^{ + }$ species versus laser fluence (Fig. \ref{fig2}, 
experimental points) shows an abrupt change in the behavior of the 
experimental data that can be treated as the change of the ejection regime. 
The first regime, corresponding to the generation of the fast population, 
exhibits a very strong non-linear dependence on fluence, $F_{0}^{10.4}$, 
that is a typical feature of the multi-particle ejection process \cite{3}. The 
second regime, starting after melting, shows near-linear Si$^{ + }$ 
abundance variation with fluence.

The expulsion of the non-thermal Si$^{ + }$ ions is attributed to 
accumulation of a positive charge on the silicon surface as a result of 
electron photoemission. This charge has been evaluated by measuring a 
compensating current variation between the Si target and ground \cite{14} during 
laser pulse. The positive charge is accumulated already at $\sim 0.001 $ 
J/cm$^{2}$ (see inset on Fig. \ref{fig1}), increasing near linearly up to a saturation 
level of $(1 - 2) \cdot 10^{13}$ elementary charges per pulse at $\sim 
 0.15 - 0.2 $ J/cm$^{2}$. Exactly at these fluences we detect the first 
non-thermal ions. Both the target charge saturation and ions ejection 
clearly indicate that Si surface reaches the critical, threshold, conditions 
corresponding to dynamical equilibrium between primary particles ejection 
(photoelectrons) and desorption of the secondary species (positively charged 
ions).
\textbf{ }
\section{Modeling}

An assumption for the electron photoemission under the UV laser irradiation, 
when photon energy exceeds work function, reads as

\begin{equation}
\label{eq1}
PE = \frac{1}{2}\alpha _{{ab}} \frac{I(x,t)}{hv}\exp \left( { - 
\frac{x}{l_{PE} }} \right)
\end{equation}

\noindent
where $I(x$,$t)$ is the laser power of a Gaussian temporal shape:

\begin{equation}
\label{eq2}
I(x,t) = \left( {1 - R} \right)\frac{2F_0 }{\tau }\sqrt {\frac{\ln 2}{\pi }}
\exp \left( { - 4\ln 2\left( {\frac{t}{\tau }} \right)^2} \right)\mbox{e}^{- \alpha_{ab} x},
\end{equation}

$\tau $ is laser pulse duration (FWHM), $\alpha_{ab}$ and $R$ are the absorption and 
reflection coefficients, $l_{PE}$ is the electron escape depth, $x$ is the 
distance from the target surface toward the bulk depth, \textit{hv }is the energy of 
laser light quantum. The expression (\ref{eq1}) implies that the laser-generated 
electrons whose momentum component normal to and in the direction of the 
surface are immediately photoemitted from the surface and below surface 
region with an exponential decreasing within the bulk \cite{5}.

\onecolumngrid

An estimate of 
the number of electrons \textit{photoemitted through a unit surface area} ($N_{PE})$ is obtained by integrating the 
photoemission term over time and space that gives:

\begin{equation}
\label{eq3}
N_{PE} = \frac{\alpha _{{ab}} (1 - R)F_0 }{2h\nu (\alpha _{{ab}} + 
l_{PE}^{ - 1} )}
\end{equation}

Under the assumption that the target is unearthed and there is no electron 
supply from the radiation-free sides of the target, the electric field 
generated on the target surface can be estimated by using the Gauss law (the 
net positive charge of the target is equal to the charge of the photoemitted 
electrons):

\begin{equation}
\label{eq4}
\left. E \right|_{x = 0} = \frac{1}{2\varepsilon \varepsilon _0 
}\int\limits_0^L {q(x)dx = \frac{eN_{PE} }{2\varepsilon \varepsilon _0 }} = 
\frac{e\alpha _{{ab}} (1 - R)F_0 }{4\varepsilon \varepsilon _0 
h\nu(\alpha _{{ab}} + l_{PE}^{ - 1} )},
\end{equation}

\noindent
where $L$ is the target width and $\varepsilon$ is dielectric permittivity of bulk silicon. 

The threshold electric field necessary to be exceeded in order to break the 
atomic bonds in crystalline silicon can be estimated through the energy 
density of the electric field, $w = \varepsilon \varepsilon_{0} E^{2} /2$. The value $W_{at} = 
\varepsilon \varepsilon_{0} E^{2}/2n$ falls at an atom in the crystal, where $n^{-1}$ is a 
volume occupied by atom. The binding energy of an atom on the target surface 
estimated from the latent heat of sublimation, $\Lambda _{sub}$ = 16115 
J/g \cite{7}, is $ \approx 4.67 $ eV. Thus, the threshold electric field is of 
order 
$E_{th} \vert _{x = 0} = \sqrt {2\Lambda_{at} n / \varepsilon \varepsilon _{0}}$ or $E_{th} \approx $ 2.65$ \cdot $10$^{10}$ V/m 
for crystalline silicon. Laser fluence above which the non-thermal ions can 
be observed is:

\begin{equation}
\label{eq5}
F_{0th} = \frac{4\varepsilon \varepsilon _0 h\nu E_{th} \left( {\alpha 
_{{ab}} + l_{PE}^{ - 1} } \right)}{e\alpha _{{ab}} (1 - R)}.
\end{equation}

With $l_{PE}\sim $ 10 {\AA} it gives $\sim 0.13 $ J/cm$^{2}$ that is in 
good agreement with the experimental observations. The excess of positive 
charge in the target for generation of the threshold electric field is 
$eN_{Ith} = 2\left. {\varepsilon \varepsilon _0 S_R E_{th} } \right|_{x = 0} 
$ or $\sim 1.8 \cdot 10^{13}$ electrons escaped from the irradiation 
spot size $S_{R}$ of 0.5 mm$^{2}$ that is in excellent agreement with the 
measured electric residue. 

A strongly charged target tends to take off the 
electrostatic stress, decreasing the electric field below its threshold 
value through ion expulsion. The number of ions thrown out from the target 
by the electrostatic force can be evaluated as

\begin{equation}
\label{eq6}
\left. {N_{\exp } } \right|_{S_R } = N_I - N_{Ith} = \frac{\alpha 
_{{ab}} (1 - R)(F_0 - F_{0th} )S_R }{2h\nu (\alpha _{{ab}} + 
l_{PE}^{ - 1} )}
\end{equation}

\noindent
or 1.4$ \cdot $10$^{14}(F_{0}-F_{0th})$ for our irradiation conditions 
(laser fluence is taken in J/cm$^{2})$. This dependence follows the general 
tendency observed experimentally (Fig. \ref{fig2}, solid line). Thus, a seeming 
change in the ablation regime from the strong non-linear to near-linear 
dependence of ion yield vs. laser fluence can be roughly described by a 
shifted linear dependence [Eq. (\ref{eq6})]. 

More sophisticated description can be reached by modeling the electron dynamics in the laser-irradiated semiconductor 
targets. Here we shall consider only the target charging effect without 
concerning the ion ejection process, both electrostatic and thermal. Our 
model is based on the continuity equations for electron and hole generation, 
including one-photon ionization, Auger recombination and the photoemission 
process according to Eq. (\ref{eq1}), and we incorporate the electric current 
$J_{y}(y=e,h) = \left|e\right|n_{y}\mu E - eD \nabla n_{y}$ written in the drift-diffusion 
approach into the equations:

\begin{equation}
\label{eq7}
\frac{\partial n_e }{\partial t} - \mu _e n_e \frac{e}{\varepsilon 
\varepsilon _0 }\left( {n_h - n_e } \right) - \mu _e E\frac{\partial n_e 
}{\partial x} - \frac{\partial }{\partial x}D_e \frac{\partial n_e 
}{\partial x} = \alpha _{{ab}} \frac{I(x,t)}{\hbar \omega } - \gamma 
n_e^2 n_h - PE,
\end{equation}

\begin{equation}
\label{eq8}
\frac{\partial n_h }{\partial t} + \mu _h n_h \frac{e}{\varepsilon 
\varepsilon _0 }(n_h - n_e ) + \mu _h E\frac{\partial n_h }{\partial x} - 
\frac{\partial }{\partial x}D_h \frac{\partial n_h }{\partial x} = \alpha 
_{{ab}} \frac{I(x,t)}{\hbar \omega } - \gamma n_e^2 n_h .
\end{equation}

Here $n, \mu, D, \gamma $ are the density, mobility and the diffusion and Auger 
recombination coefficients, respectively; indices $e$ and $h$ refer to the 
electrons and holes. The diffusion coefficients are calculated as $D_{e} = 
k_{B}T_{e}\mu_{e}/e$ and $D_{h}=k_{B}T_{e}\mu_{h}/e$ with the carrier 
temperature $T_{e}$. The photoemission and diffusion lead to charge separation 
in the target that results in the electric field generation described by the 
Poisson equation with the boundary condition according to the Gauss law. The 
energy equations for the electron and lattice subsystems \cite{15} degenerate 
into an only equation at fairly long laser pulses of ns time scale \cite{16,17}

\begin{equation}
\label{eq9}
c_p \rho \left( {\frac{\partial T}{\partial t} - 
u(t)\frac{\partial T}{\partial x}} \right) = \frac{\partial 
}{\partial x}\lambda \frac{\partial T}{\partial x} + 
\alpha _{{ab}} (1 - R) I_0 (t)\exp ( - \alpha _{{ab}} x).
\end{equation}

\twocolumngrid

The vaporization rate $u(t)$ is defined under the assumption that the flow of 
vaporized material from the surface follows the Hertz-Knudsen equation and 
the vapor pressure above the vaporized surface can be estimated with the 
Clausius-Clapeyron equation:

\begin{equation}
\label{eq10}
u(t) = (1 - \beta )\frac{p_b }{\rho }\left( {\frac{m}{2\pi kT_s }} 
\right)^{1 / 2}\exp \left[ {\frac{L}{k}\left( {\frac{1}{T_b } - \frac{1}{T_s 
}} \right)} \right],
\end{equation}

Thermodynamic and optical properties of silicon (specific heat $c_{p}$, 
thermal conductivity $\lambda $, latent heat of vaporization $L$, boiling temperature 
$T_{b}$, absorption coefficient $\alpha_{ab}$, and reflection coefficient $R$) were 
taken from Ref. \cite{9}. The initial and boundary conditions are

\begin{equation}
\label{eq11}
T(0,x)=T_{0},~~T(t,0) = T_{s}(t),~~\lambda \left. {\frac{\partial T}{\partial x}} \right|_{x = 0} = Lu(t)
\end{equation}

\noindent
with $T_{0}$ being the initial temperature uniform across the target. The 
target was divided into an irregular grid, dense in the absorption region 
(cells of 5 {\AA}) and rarefying toward the target depth. We used an 
explicit scheme for solving the described system of equations. The time step 
was selected empirically to satisfy the numerical scheme stability and the 
approximation of the original equations. 

\section{Results and discussion}

\begin{figure}[!]
\resizebox{8cm}{!}{\includegraphics{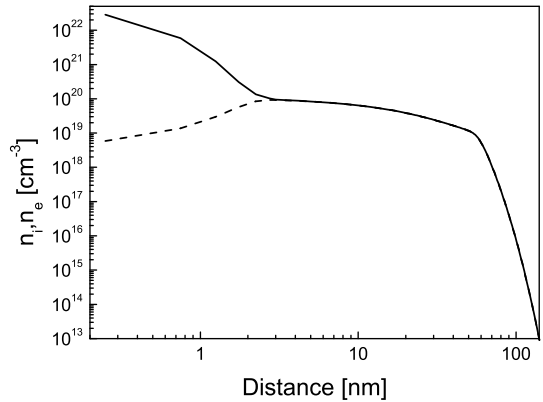}} \caption{\label{fig3}
Calculated spatial profiles of the electron (dashed line) 
and ion densities (solid line) in the Si target irradiated with 0.2 
J/cm$^{2}$.}
\end{figure}

\begin{figure}[!]
\resizebox{8cm}{!}{\includegraphics{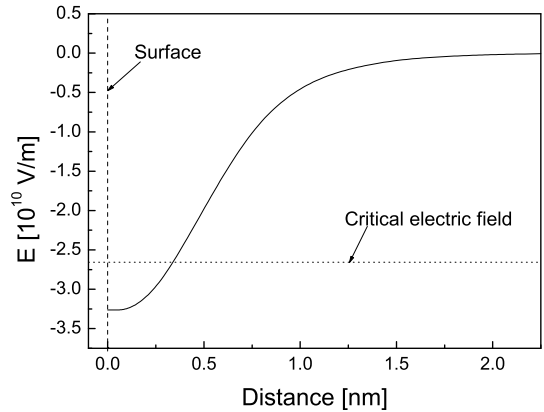}} \caption{\label{fig4}
Spatial distribution of the electric field in the Si target 
($F_{0}$ = 0.2 J/cm$^{2})$.}
\end{figure}

The modeling results are presented in Figs. \ref{fig3}, \ref{fig4}. To the end of the laser 
pulse, the surface charge reaches almost 60{\%}, an unexpected value for 
silicon irradiated by low laser fluence. The densities of the electrons and 
holes are presented in Fig. \ref{fig3}. Considerable charging takes place in a 
surface zone of order of 2 nm wide ($\sim 5 - 6$ atomic monolayers). It is 
generally agreed that, because of efficient Auger recombination, a silicon 
target can not be ionized to a high ionization degree at moderate laser 
fluences. That is a major argument against the Coulomb explosion mechanism. 
Our modeling shows that a thin surface layer of the target reaches a high 
ionization degree, but therewith this layer occurs to be strongly depleted 
of electrons. Extremely high density of positive charge is gained in a few 
monolayers, implying the strong repulsion force between the ions. First of 
all, adatoms, which have lower binding with surface, have to be ejected from 
the surface, giving way to the generation of new adatoms. The process of ion 
expulsion will take place till all the layer of high positive charge is 
exploded or till the macroscopic electric field is reduced below the 
threshold value. Figure \ref{fig4} illustrates that, at 0.2 J/cm$^{2}$, the 
macroscopic electric field exceeds the threshold value, as it was predicted 
analytically. The values of charging and the electric field in two external 
monolayers are so high, that their electrostatic explosion is inevitable. 

Let us consider the origin of high surface charging of a silicon target in 
detail. What is happening when a target subjected to laser irradiation loses 
the electrons due to photoemission? The dielectric breakdown takes place 
within a skin layer with an exponential decay toward the target depth. 
Because of photoemission, the target quasi-neutrality is broken. According 
to the Gauss law, the target, which starts to behave as a metal, aspires to 
accumulate the excess positive charge in its surface. The electric field in 
this thin surface layer is ``negative'' (directed to vacuum) that means that 
the non-emitted electrons are dragged toward the target depth to a region 
with a lower field. Thus, the following scenario of the electrostatic 
mechanism for both ion desorption and ablation from the semiconductor 
targets (that may be a case also for dielectrics) can be proposed step by 
step: (i) target material breakdown and photoemission under laser 
irradiation; (ii) generation of the electric field which drags the electrons 
away from the surface layer; (iii) continuing surface ionization together with 
suppression of Auger recombination; (iv) as a result, ionization degree up to 
100{\%} in a thin surface layer; (v) electrostatic disintegration of this 
layer. \textit{The most important consequence is that, in the charged outer layer, one-photon ionization is continuing during the laser pulse, whereas the Auger recombination is strongly suppressed because of lack of electrons.} 

\section{Conclusions}

Mechanisms of desorption and ablation of silicon under irradiation by the 
nanosecond laser pulses in a wide range of laser fluences have been studied 
both experimentally and theoretically. At low laser fluence the desorption 
flux is mostly composed of high-energy ions with a narrow energy 
distribution that points to a non-thermal mechanism of their generation. 
Theoretical analysis has attributed the high-energetic ions to a strong 
electric field generated in an exterior target layer due to photoemission 
that causes Coulomb explosion. More sophisticated two-temperature modeling, 
which takes into account electronic ion emission and the aspects of thermal 
vaporization, is under the progress.

\end{document}